# Electric field control of nonvolatile four-state magnetization at room temperature

Sae Hwan Chun,[1][*] Yi Sheng Chai,[1][†] Byung-Gu Jeon,[1] Hyung Joon Kim,[1] Yoon Seok Oh,[1] Ingyu Kim,[1] Hanbit Kim,[1] Byeong Jo Jeon,[1] So Young Haam,[1] Ju-Young Park,[1] Suk Ho Lee,[1] Jae-Ho Chung,[2] Jae-Hoon Park,[3] and Kee Hoon Kim[1]

[1]*CeNSCMR, Department of Physics and Astronomy, Seoul National University, Seoul 151-747, Korea*

[2]*Department of Physics, Korea University, Seoul 136-713, Korea*

[3]*Department of Physics and Division of Advanced Materials Science, POSTECH, Pohang 790-784, Korea*

*E-mail: pokchun@snu.ac.kr

†E-mail: hawkchai@gmail.com

**Abstract**

We find the realization of large converse magnetoelectric (ME) effects at room temperature in a multiferroic hexaferrite $Ba_{0.52}Sr_{2.48}Co_2Fe_{24}O_{41}$ single crystal, in which rapid change of electric polarization in low magnetic fields (about 5 mT) is coined to a large ME susceptibility of 3200 ps/m. The modulation of magnetization then reaches up to 0.62 $\mu_B$/f.u. in an electric field of 1.14 MV/m. We find further that four ME states induced by different ME poling exhibit unique, nonvolatile magnetization versus electric field curves, which can be approximately described by an effective free energy with a distinct set of ME coefficients.

The capability of the electrical control of magnetization at room temperature becomes increasingly important for many contemporary or next-generation devices such as a multi-bit memory [1] or a novel spintronics apparatus [2]. Multiferroics, in which magnetism and ferroelectricity coexist and couple to each other, could be the most plausible candidates to expect such a converse ME effect via their gigantic ME coupling that generally covers both linear and nonlinear effects [3-8]. Indeed, the vibrant multiferroics research during the past decade has led to significant advances such as the discovery of $BiFeO_3$ with large electric polarization $P$ [9] and numerous magnetic ferroelectrics allowing sensitive $P$ control by a magnetic field $H$ [3,4,10-12]. However, when the general magnetoelectric susceptibility (MES), $dP/dH$, is evaluated at room temperature, a $BiFeO_3$ single crystal exhibits the maximum MES of ~55 ps/m (at $\mu_0 H \leq 16$ T) [13]. Furthermore, most of the magnetic ferroelectrics indeed show the giant ME coupling far below room temperature. Although a recent discovery of the room-temperature ME effect in a polycrystalline $Sr_3Co_2Fe_{24}O_{41}$ has greatly increased the operation temperature, its maximum MES of 250 ps/m [12] is still not enough to expect a significant converse effect. Moreover, even the well-known magnetoelectrics such as $Cr_2O_3$ [14] and $GaFeO_3$ [15] did not show substantial modulation of bulk magnetization $M$ by an electric field $E$ at room temperature, mainly due to their weak ME coupling strengths ($dP/dH \leq 5$ ps/m). Therefore, the large room-temperature modulation of $M$ by $E$ still remains as one of great challenges that should be realized in multiferroic or magnetoelectric compounds [16-18].

To overcome the long-standing challenge, we focus here the $Co_2Z$-type hexaferrite $(Ba,Sr)_3Co_2Fe_{24}O_{41}$, which consists of a series of tetrahedral and octahedral Fe/Co layers stacked along the $c$ axis ($\parallel [001]$) [Fig. 1(a)]. $Sr_3Co_2Fe_{24}O_{41}$ is known to have a collinear ferrimagnetic ordering below 670 K, in which the net magnetic moments in alternating large ($L$) and small ($S$) blocks are oriented parallel to the $c$ axis but antiparallel to each other ($\mu_L > \mu_S$). These moments rotate away from the $c$ axis towards the $ab$ plane below 500 K. The ME effect appears below 400

K where a transverse conical magnetic order appears with the spin propagation vector $\vec{k}_0$ = (0, 0, 1) [Fig. 1(b)] [19]. As the room-temperature ME effect was realized only in a polycrystalline $Sr_3Co_2Fe_{24}O_{41}$, we have then tried to grow single crystals of $(Ba,Sr)_3Co_2Fe_{24}O_{41}$ to increase the MES by either controlling directions of electric/magnetic dipoles or by tuning the chemical pressure. In this Letter, we report the successful single crystal growth of $Ba_{3-x}Sr_xCo_2Fe_{24}O_{41}$ in a broad $x$ range. In a single crystal showing the highest MES among the series, we discover significant converse ME effects at room temperature. Furthermore, we show that the compound can save four distinct $M(E)$ curves in a non-volatile way via independent ME writing conditions.

$Ba_{3-x}Sr_xCo_2Fe_{24}O_{41}$ ($2.40 \leq x \leq 3.00$) single crystals were grown by the flux method. Among at least four different compositions including $Sr_3Co_2Fe_{24}O_{41}$, a $Ba_{0.52}Sr_{2.48}Co_2Fe_{24}O_{41}$ single crystal showed the highest MES so that it was selected for the present study [20]. Superior to other hexaferrites [10, 11, 21-23], the $Ba_{0.52}Sr_{2.48}Co_2Fe_{24}O_{41}$ crystal exhibits quite high resistivity upon heat treatment [22] ($2.5 \times 10^9 \, \Omega$ cm at 300 K and $1.0 \times 10^7 \, \Omega$ cm at 380 K), which allows us to measure reliably MES up to 420 K and thus to extract the corresponding $P(H)$ curves. The $M(T)$ curves measured at either $\mu_0H$ = 50 or 500 mT [Fig. 1(c)] further indicate that the ferrimagnetic order starts to develop at 680 K and magnetic easy axis changes at 490 K, followed by the stabilization of the transverse conical spin order below $T_{con}$ = 410 K [19]. Our independent neutron diffraction study on the $Ba_{0.52}Sr_{2.48}Co_2Fe_{24}O_{41}$ crystal revealed the development of (0, 0, $l$) ($l$ = odd) peaks below $T_{con}$, being consistent with the case of a polycrystalline $Sr_3Co_2Fe_{24}O_{41}$ [19]. Moreover, the neutron diffraction data could be successfully fitted by the transverse conical spin order as depicted in Fig. 1(b).

As demonstrated in Fig. 1(d), nontrivial $P$ ($\| [100]$) is induced under a transverse $H$ ($\| [120]$) in a wide temperature range below $T_{con}$ = 410 K, yielding maximum $\Delta P (\equiv P(\mu_0H) - P(\mu_0H_p \equiv 2 \, T))$ values of 32 $\mu C/m^2$ at 305 K and 77 $\mu C/m^2$ at 10 K. Note that $P$ rapidly increases in the 5-50 mT range, in which the magnetic hysteresis curves also show sharp, quasi-linear changes [Fig. 1(e)]. This observation suggests that the spin cone axes are probably disordered in zero or small fields, whereas they are stabilized into transverse cones at $\mu_0 H \geq 50$ mT [Fig. 1(b)]. According to the spin-current model [24-26], where $\vec{P} \propto \sum \vec{k}_0 \times (\vec{\mu}_L \times \vec{\mu}_S)$ and $\vec{k}_0$ ($\parallel$ [001]) is the spin propagation vector, the transverse cones depicted in Fig. 1(b) allow a finite $P$ ($\parallel$ [100]). Therefore, alignment of the spin rotation axis toward the $H$ direction will greatly enhance $P$. In a higher $H$ region (> 1 T), in contrast, the transverse conical component will be reduced at the expense of the collinear ferrimagnetic one, eventually suppressing $P$. On the other hand, a magnified $M(H)$ curve at 305 K reveals a small hysteresis with a remanent $M = 0.34$ $\mu_B$/f.u. and a small coercive field $\mu_0 H_c \approx 0.3$ mT [Fig. 1(e)], consistent with a soft ferrimagnet-like behavior in the $H$-controlled crossover between the disordered and ordered transverse cone states. The observation of the remanent $M$ indicates that the transverse cone axes might not be completely disordered to contribute also to non-vanishing $\Delta P$ even in zero $H$-bias.

Direct MES measurements with a very small ac field $\mu_0 H_{ac} = 0.02$ mT by use of a special ME susceptometer developed for detecting quite small charge oscillation down to $\sim 10^{-17}$ C [27], indeed supported the possible existence of a non-vanishing $\Delta P$ involving the remanent $M$. Figures 2(a) and 2(b) show the MES data obtained after a ME poling procedure, in which $H$ was decreased from $+\mu_0 H_p \equiv 2$ T to 1.2 T under application of $+E_p \equiv 230$ kV/m or $-E_p$ and the electric field was subsequently turned off, followed by an electrical shorting of the electrodes. Upon decreasing $H$ from 1.2 T, the MES value for $+E_p$ increases to reach a maximum of 3200 ps/m at $\mu_0 H = 10.5$ mT, and then it decreases almost linearly to have finite intercepts in both $H$ (0.25 mT) and d$P$/d$H$ (-149 ps/m) axes [Fig. 2(b)]. Eventually it reaches a minimum of -2500 ps/m at -10.5 mT. We note that 3200 ps/m is the highest MES value ever observed in magnetoelectric or multiferroic compounds at room temperature [12-15]. Thus, the $\Delta P(H)$ curve, obtained by integrating d$P$/d$H$ with $H$ for $+E_p$, shows the steepest slope at 10.5 mT [Fig. 2(c)] and the minimum ($\Delta P_{min} \approx 0$) at 0.25 mT [Fig. 2(d)]. At $H = 0$, we obtained a finite $\Delta P = 0.016$ μC/m$^2$. For -$E_p$, the signs of MES as well as of $\Delta P$ are reversed to leave once again a finite $\Delta P = -0.016$ μC/m$^2$ at $H = 0$. When the same experiment was repeated upon increasing $H$ after the similar ME poling process starting at -$H_p$, the $\Delta P(H)$ curve consistently shows a similar sign reversal with change of electric poling direction. These observations suggest that the spin helicity, which determines the sign of the induced $P$ according to the spin-current model, can be controlled by the electric poling direction. The finite offset in the $\Delta P$ curve at zero $H$-bias is likely associated with the memory effects in the spin helicity and the in-plane magnetization coming from the ME poling [28].

In order to utilize the memory effects in the $M$ control, we first employed the ME poling (+$H_p$ & +$E_p$), termed "*state* **0**", in which $H$ was decreased from +$\mu_0 H_p \equiv 2$ T to 1.2 T under application of +$E_p \equiv 230$ kV/m and the electric field was subsequently turned off. $M$ (|| [120]) was then measured at a fixed, bias $H$ (|| $M$) while applied $E$ (|| [100]) is swept slowly in time at a frequency of 0.02 Hz. We found that $M$ varies significantly with the $E$ sweep in a broad $H$-bias region below 2 T. The largest modulation was indeed observed at a bias $\mu_0 H = 10.5$ mT, at which the MES value becomes the maximum. Figure 2(e) shows that the $M$ modulation at 10.5 mT reaches as large as |$\Delta M \equiv M(E) - M(0)$| = 0.62 μ$_B$/f.u., which corresponds to the relative change of |$\Delta M$|/$M(E=0) \approx 6.0$ %. Moreover, the modulation was reproducible for many $E$-cycles. When the sign of $H$-bias was reversed (-10.5 mT), we obtained somewhat reduced |$\Delta M$| = 0.45 μ$_B$/f.u. and |$\Delta M$|/$M(E=0) \approx 4.6$ % [Fig. 2(f)]. These results are consistent with the smaller absolute MES value at the negative $H$ region. In addition, as a natural consequence of the high resistivity in this system, the applied $E$ does not generate any significant leakage current, and thus its $M$ modulation brings only extremely small energy dissipation. Therefore, the current result demonstrates non-dissipative, repeatable large $M$ control by $E$ at room temperature, significantly larger than the highest repeatable modulation known to date at 15 K (|$\Delta M$| < 3 × 10$^{-3}$ μ$_B$/f.u.) [16].

The converse ME effect could be further controlled by ME poling and $H$-bias conditions. The $M(E)$ curves shown in Figs. 2(e) and 2(f) indeed show asymmetric line shapes, which can be decomposed into linear and quadratic terms of $E$ that are strongly dependent upon the value of $H$-bias. For the bias $\mu_0 H = 10.5$ mT (-10.5 mT), the $M(E)$ curve shows a positive (negative) linear slope and a negative (positive) quadratic curvature. Remarkably, we found that such asymmetric $M$ modulation is sustained even without the $H$-bias after the initial ME poling. The $M(E)$ curve obtained after applying the "*state* **0**" [Fig. 3] shows a larger $M$ modulation in a negative $E$ region except for a hysteresis near zero $E$, thereby giving a characteristic, asymmetric parabola shape. We note that the asymmetric line shape was always reproducible regardless of the small hysteresis, which was slightly dependent on the ME poling history [20]. When the ME poling is changed into $+H_p$ & $-E_p$ (*state* **1**), the $M(E)$ curve exhibits clearly different parabola shape with a larger $M$ modulation in a positive $E$ region. When the same experiment was repeated with $-H_p$ & $+E_p$ (*state* **2**) or $-H_p$ & $-E_p$ (*state* **3**), in which $H$ was decreased from $-\mu_0 H_p \equiv -2$ T to $-1.2$ T under application of $+E_p \equiv 230$ kV/m or $-E_p$ and the electric field was subsequently turned off, the $M(E)$ loops display their own characteristic line shapes, distinguished from the loops for the *state* **0** or **1**. As the result, the four ME poling conditions produce distinct four $M(E)$ loops. Moreover, we also found that all the $M(E)$ loops are persistently repeatable for many $E$-cycles, proving that the ME information is non-volatile and cannot be erased by the application of $E$ [20]. These results suggest a unique opportunity to store nonvolatile, two-bit information in this single crystal; not only these four states can be written but also can be read by the independent, four $M$ values under a small $E$-bias. Moreover, under finite $\pm E$-biases, each asymmetric $M(E)$ curve displays two independent $M$ values. This observation also suggests yet another possibility to store eight different states of $M$ or four-bit information into this material under a finite $E$.

The $M(E)$ curves in a low $H$-bias suggest that each curve can be effectively described by a free energy $F(E,H)$ [5],

$$F(E,H) = -PE - \mu_0 MH + \frac{1}{2}\varepsilon_0\varepsilon E^2 + \frac{1}{2}\mu_0\mu H^2 + \alpha EH + \frac{1}{2}\beta EH^2 + \gamma E^2 H + \frac{1}{2}\pi E^2 H^2 + ... \quad (1),$$

where $\varepsilon$ and $\mu$ ($\varepsilon_0$ and $\mu_0$) are dielectric permittivity and magnetic permeability of a material (vacuum) and $\alpha$, $\beta$, $\gamma$ and $\pi$ are the linear and higher order ME coefficients. By minimizing $F$ with respect to $H$, the modulations of $M$ with $E$ and $H$ can be described as:

$$\mu_0 \Delta M(E,H) \equiv \mu_0\{M(E,H) - M(0,H)\} = (\alpha + \beta H)E + (\gamma + \pi H)E^2 + ... \quad (2),$$

which indicates that the derived $\Delta M(E, H)$ can be approximated by a sum of linear and quadratic $E$ terms in a low $E$ region, being consistent with observations of an asymmetric line shape in the $M(E)$ curves [Figs. 2 and 3]. Eq. (2) further suggests that the coefficients of these linear and quadratic $E$ terms should be linearly dependent upon the $H$-bias. To check this for each ME state [Fig. 4(a)-(d)], we fitted experimental $\Delta M(E, H)$ vs. $E$ curves at different $H$-biases ($|\mu_0 H| \leq 10.5$ mT) with Eq. (2).

Figures 4(e) and 4(f) summarize thus determined coefficients of linear and quadratic $E$ terms ($C_1$ and $C_2$) vs. $H$-bias, respectively. Both coefficients are indeed approximately linear in $H$ especially near $H=0$, and we fitted the curves in Fig. 4(e) (Fig. 4(f)) with a functional form of $\alpha + \beta H$ ($\gamma + \pi H$) to determine the ME coefficients $\alpha$, $\beta$, $\gamma$ and $\pi$ for each ME state. We note that the observation of non-zero $\alpha$, $\beta$, $\gamma$ and $\pi$ values is consistent with the expected ME coefficients in a transverse conical spin ordering, for which its magnetic point group 2' is expected in a finite $H$-experiment [19,20]. We further note that the resultant ME coefficients have almost the same magnitudes but characteristic signs for different ME states. This directly proves that the written ME states can be described by a distinct set of ME coefficients. Moreover, both direct ME and converse ME effects at 305 K can appear in accordance with those stored ME states. It is further inferred that the microscopic parameters such as spin helicity, cone shape, and net spin direction can be also uniquely determined for each ME state, which is characterized by a unique set of the ME coefficients. Therefore, we expect that further understanding on the relationship between the microscopic spin configuration and those ME coefficients are likely to help optimizing the converse ME effects at room temperature.

The hysteretic behavior observed in Fig. 3 should be another important phenomena that are worthy of further investigations. We found out that the hysteretic part, albeit small, was somewhat dependent upon the history of the applied $E$-bias and its magnitude. For example, it was reduced upon decreasing maximum $E$-value or applying $E$-bias down to zero $H$ region [20]. It seems thus associated with the partial switching of magnetoelectric domains under application of rather a high electric field, which can be sensitive to the history and the magnitude of the applied electric field in low $H$ regions. Although the hysteresis turns out to be rather small in this compound, maximizing such hysteretic effects would be desirable for realizing distinct $M$ values even without $E$-bias.

In summary, we demonstrated the large bulk magnetization modulation by an electric field at room temperature in the hexaferrite $Ba_{0.52}Sr_{2.48}Co_2Fe_{24}O_{41}$ crystal. Four independent magnetization vs. electric field curves, generated by the ME poling processes, represent four different ME states and can be described by a unique set of ME coefficients. Our result provides a proof of feasibility for utilizing the converse ME effect for a multi-bit memory, and thus may pave a pathway into development of a practical ME device at room temperature.

This work is supported by CRI (2010-0018300), BSR (2009-0083512) programs and the Fundamental R&D program for Core Technology of Materials. J.-H.C. is supported by the BAERI (2010-0017423) and the Nuclear R&D Program (2010-0018369). J.-H.P. is supported by CRI (2009-0081576), WCU (R31-2008-000-10059-0) and LFRIR (2010-00471) programs.

[27]   H. Ryu *et al*., Appl. Phys. Lett. **89**, 102907 (2006); Y. S. Oh *et al*., Appl. Phys. Lett. **97**, 052902 (2010); K.-T. Ko *et al*, Nat. Commun. **2**, 567 (2011); the resolutions of MES and $\Delta P$ in our sample were estimated to be about 0.1 ps/m and $10^{-5}$ $\mu C/m^2$, respectively.

[28]   Ferroelectricity could not be confirmed based on a *P-E* hysteresis loop at 305 K, while the remanent $\Delta P$ still showed a sign change upon applying a post-electric field poling [20].
10

**Figure captions**

**FIG. 1** (color online) (a) The Z-type hexaferrite structure in the hexagonal setting. (b) Schematic illustration of transverse conical spin states realized under a finite $H$. $\mu_L$ and $\mu_S$ (arrows) refer to the effective magnetic moment in the $L$ and $S$ blocks. (c) Temperature dependence of $M \parallel [120]$ and $\parallel [001]$ measured at $\mu_0 H = 50$ (dashed) and 500 mT (solid) after the field cooling. (d) $\Delta P(H) (\equiv P(H) - P(H_p))$ curves after ME poling with $+\mu_0 H_p \equiv 2$ T ($\parallel [120]$) and $+E_p \equiv 230$ kV/m ($\parallel [100]$). (e) $M(H)$ curves at 10 and 305 K. The inset shows the magnified curve at 305 K.

**FIG. 2** (color online) (a) $H$-dependence of MES for $+E_p$ and $-E_p$ cases at 305 K and (b) the same curve magnified near $H = 0$. (c) $\Delta P(H)$ curves and (d) the same curves enlarged near $H = 0$. Arrows indicate $H$-decreasing or increasing runs. (e), (f) $M(E)$ curves under bias $\mu_0 H = 10.5$ & -10.5 mT, respectively, after the ME poling with $+H_p$ and $+E_p$. Each $M(E)$ curve can be fitted with the curve (solid line) composed of linear (dashed line) and quadratic components (dash-dotted line).

**FIG. 3** (color online) The $M(E)$ curves at zero $H$-bias obtained after applying four different ME poling (*states* **0**, **1**, **2**, and **3**) as indicated in the inset.

**FIG. 4** (color online) (a)-(d) $\Delta M(E)$ curves under different $H$-bias for the four different ME poling conditions (*states* **0-3** in Fig. 3). (e), (f) The coefficients $C_1$ and $C_2$ vs. $H$-bias for *states* **0-3** upon fitting the experimental $\Delta M$ curves in (a)-(d) with $\Delta M = C_1 E + C_2 E^2$.

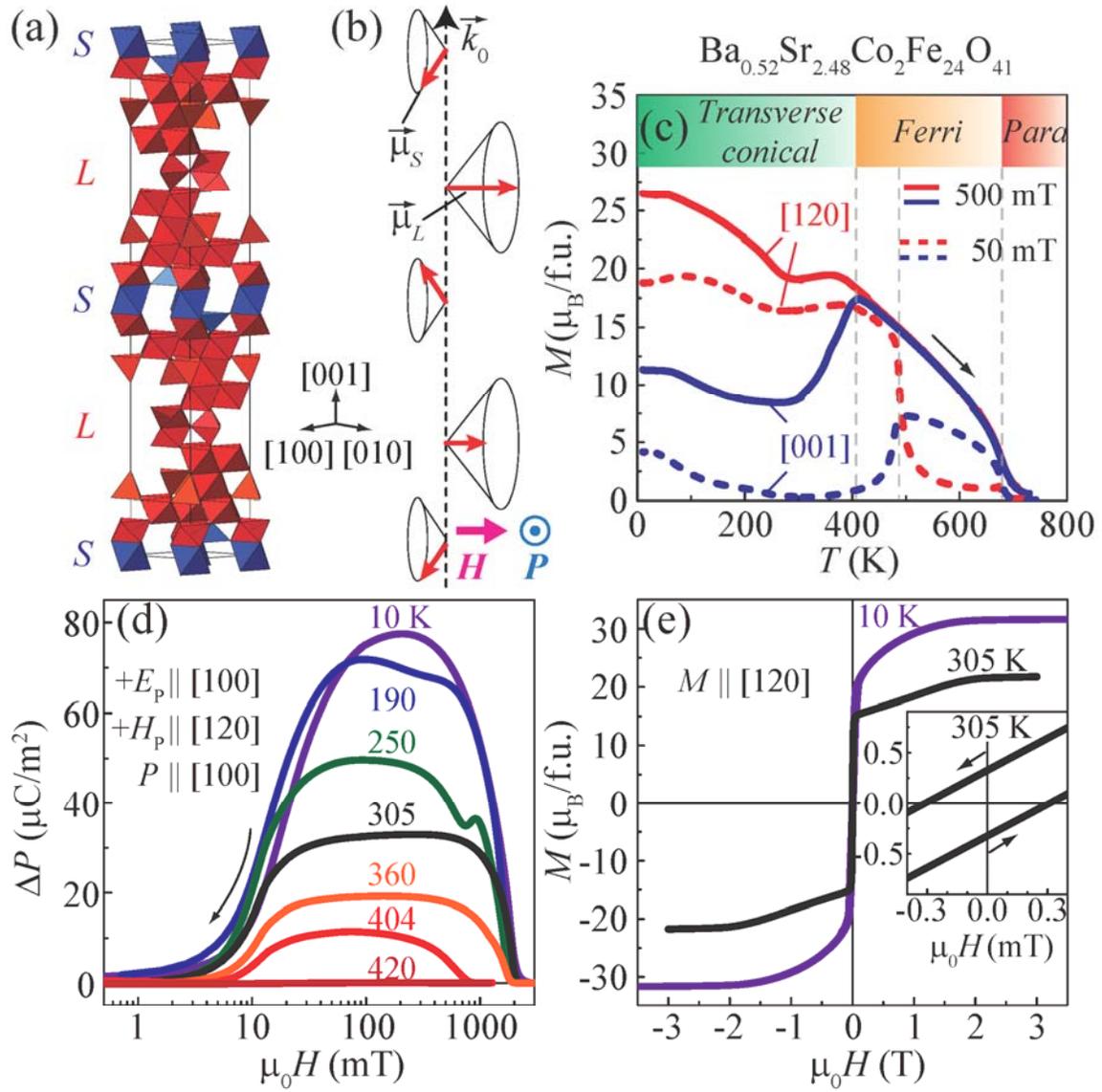

FIG. 1   S. H. Chun *et al.*

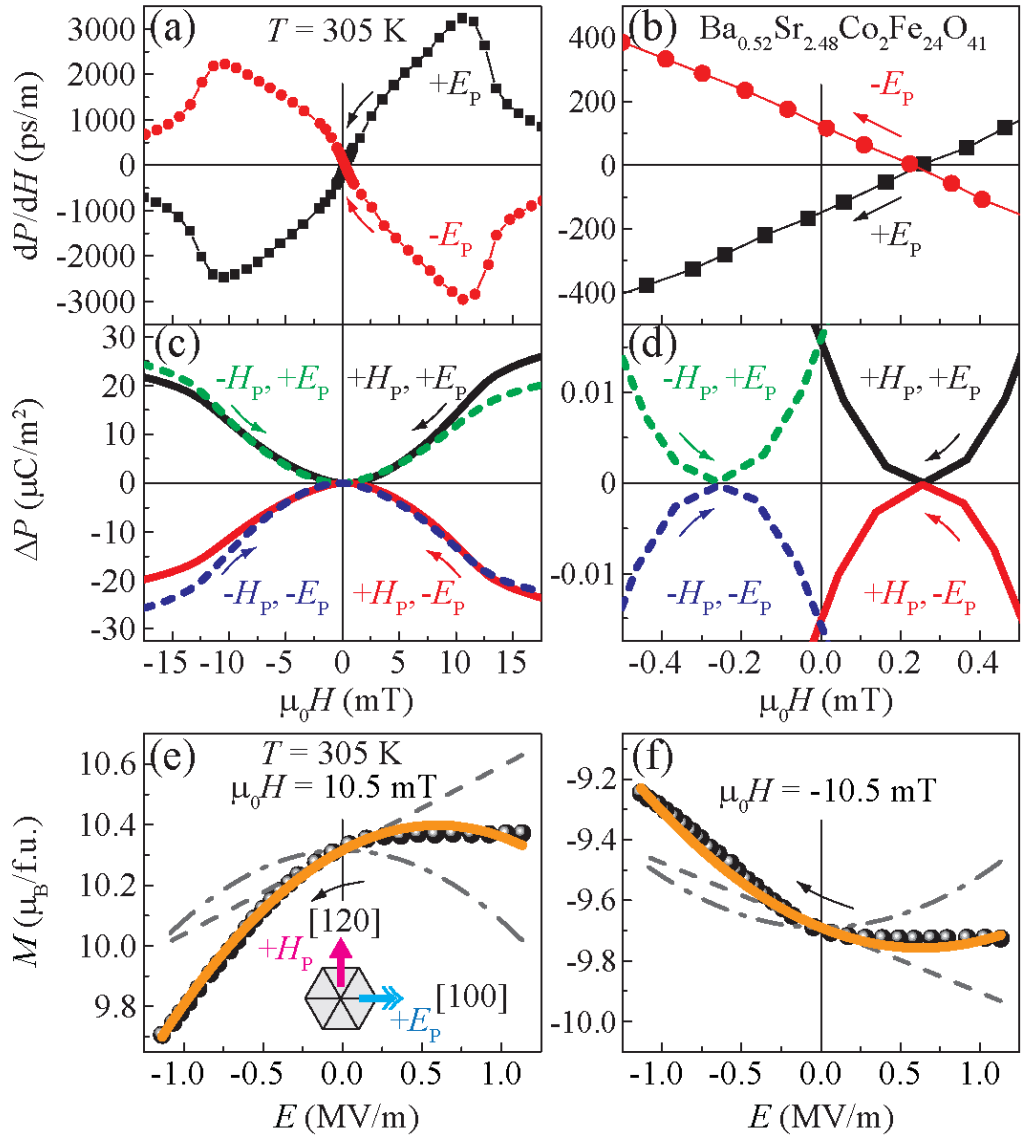

FIG. 2   S. H. Chun *et al.*

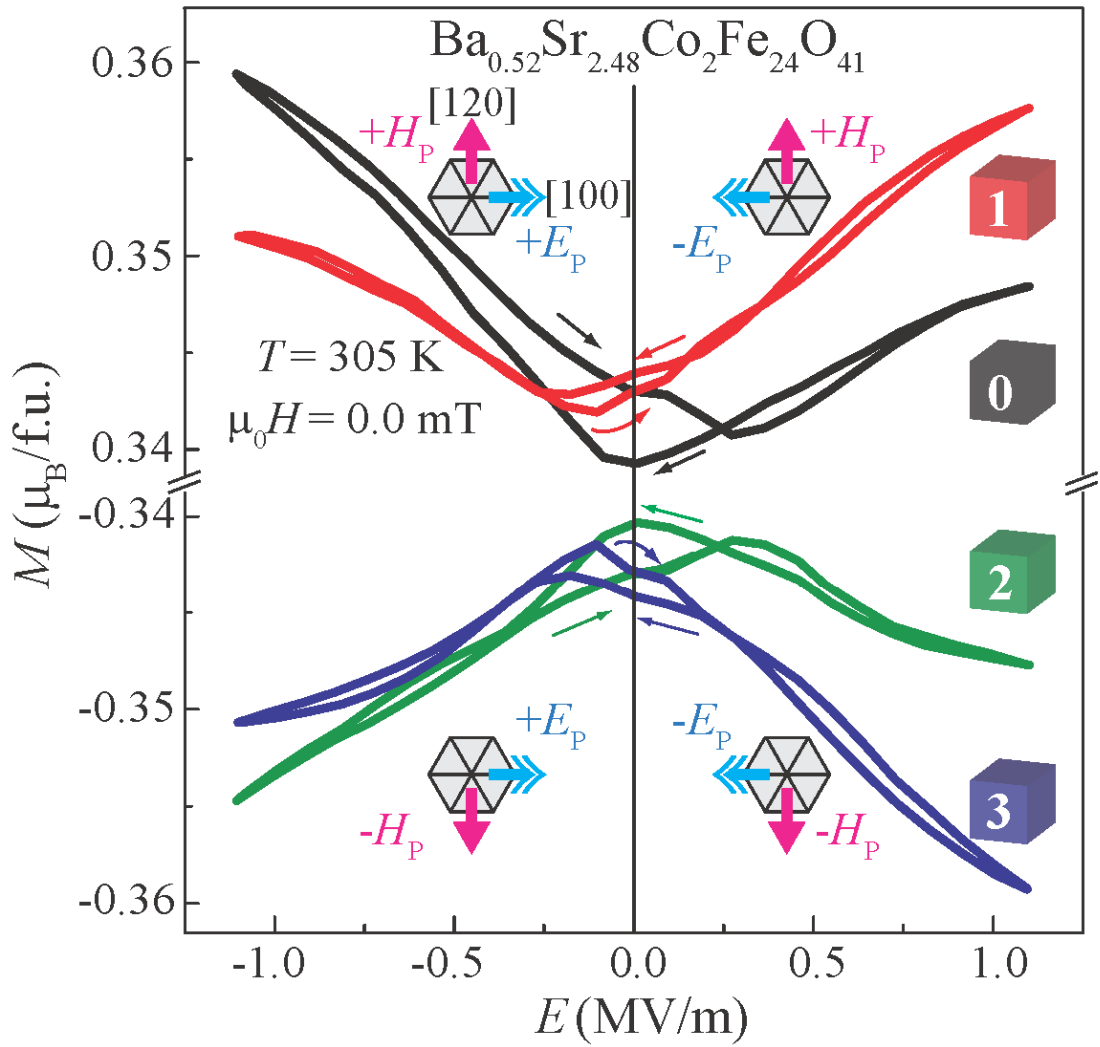

FIG. 3  S. H. Chun *et al.*

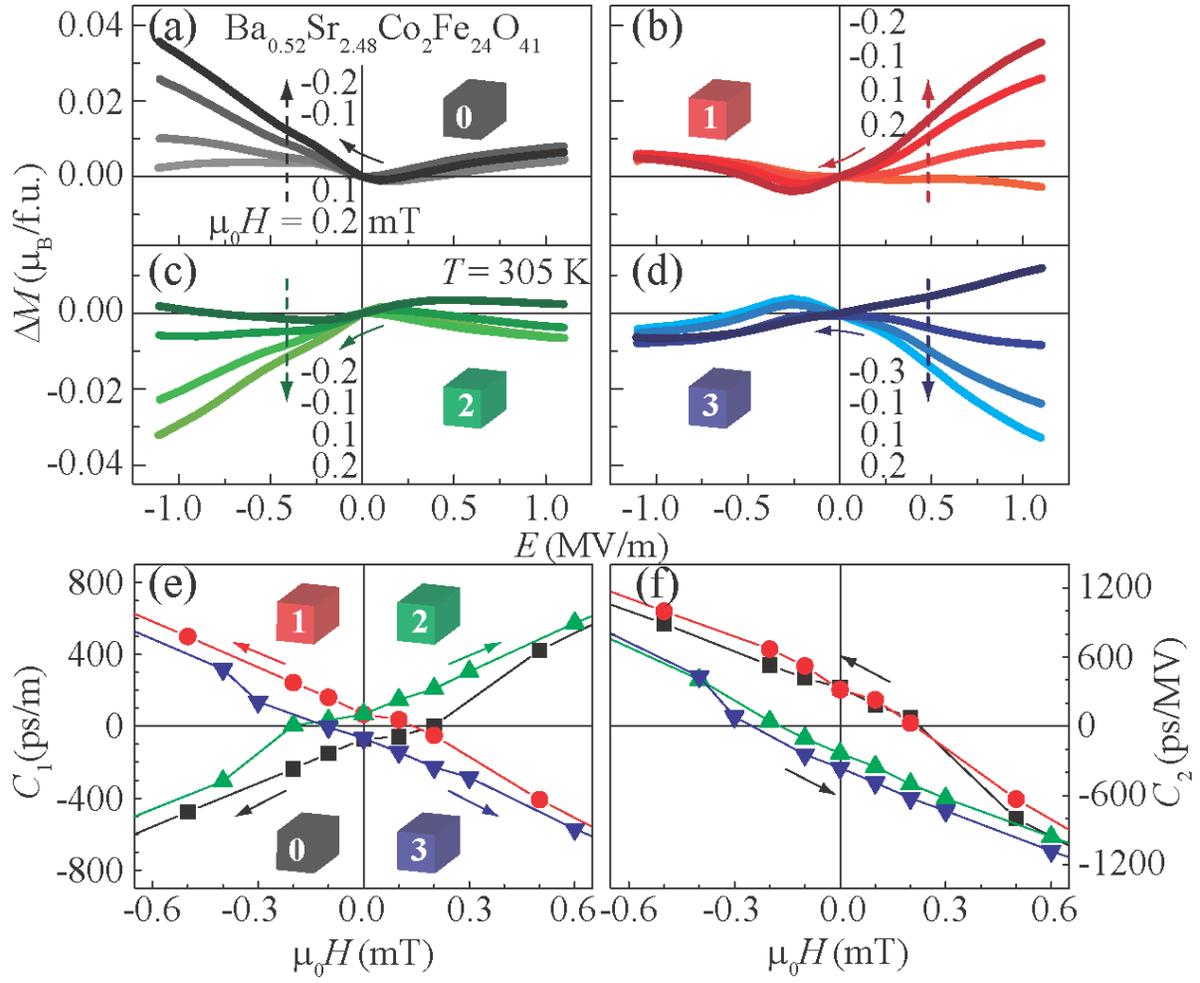

FIG. 4   S. H. Chun *et al.*

Supplemental Material for

**Electric field control of nonvolatile four-state magnetization at room temperature**

Sae Hwan Chun,[1*] Yi Sheng Chai,[1†] Byung-Gu Jeon,[1] Hyung Joon Kim,[1] Yoon Seok Oh,[1] Ingyu Kim,[1] Hanbit Kim,[1] Byeong Jo Jeon,[1] So Young Haam,[1] Ju-Young Park,[1] Suk Ho Lee,[1] Jae-Ho Chung,[2] Jae-Hoon Park,[3] and Kee Hoon Kim[1]

[1]*CeNSCMR, Department of Physics and Astronomy, Seoul National University, Seoul 151-747, Korea*

[2]*Department of Physics, Korea University, Seoul 136-713, Korea*

[3]*Department of Physics and Division of Advanced Materials Science, POSTECH, Pohang 790-784, Korea*

*Email: pokchun@snu.ac.kr

†E-mail: hawkchai@gmail.com

## 1. Sample preparation

Ba$_{3-x}$Sr$_x$Co$_2$Fe$_{24}$O$_{41}$ (2.40 ≤ $x$ ≤ 3.00) single crystals with the Z-type hexaferrite structure were grown from the Na$_2$O-Fe$_2$O$_3$ flux. The chemicals were mixed with the molar ratio of BaCO$_3$ : SrCO$_3$ : CoO : Fe$_2$O$_3$ : Na$_2$O = 19.69 $x'$ : 19.69 (1 - $x'$) : 19.69 : 53.61 : 7.01 and melted at 1420 °C in a Pt crucible, followed by several thermal cycles. The $x'$ was chosen for each target $x$ following Ref. S1. The single crystals of (Ba,Sr)$_3$Co$_2$Fe$_{24}$O$_{41}$ (Z-type) and (Ba,Sr)$_2$Co$_2$Fe$_{12}$O$_{22}$ (Y-type) turned out to be grown in the same batch. The Z-type crystals were collected by checking the $c$ axis lattice parameter from the X-ray diffraction study, and heat-treated to remove oxygen vacancy at 900 °C under flowing O$_2$ for 8 days [Ref. S2]. The final product was confirmed to be a single phase using X-ray diffraction on a crushed powder form [Fig. S1]. Chemical compositions of the crystals were determined by electron-probe-mirco-analysis (EPMA). The composition was homogeneous over the sample surface within the resolution of the EPMA, and the errors in the determined chemical ratio were about 10 %.

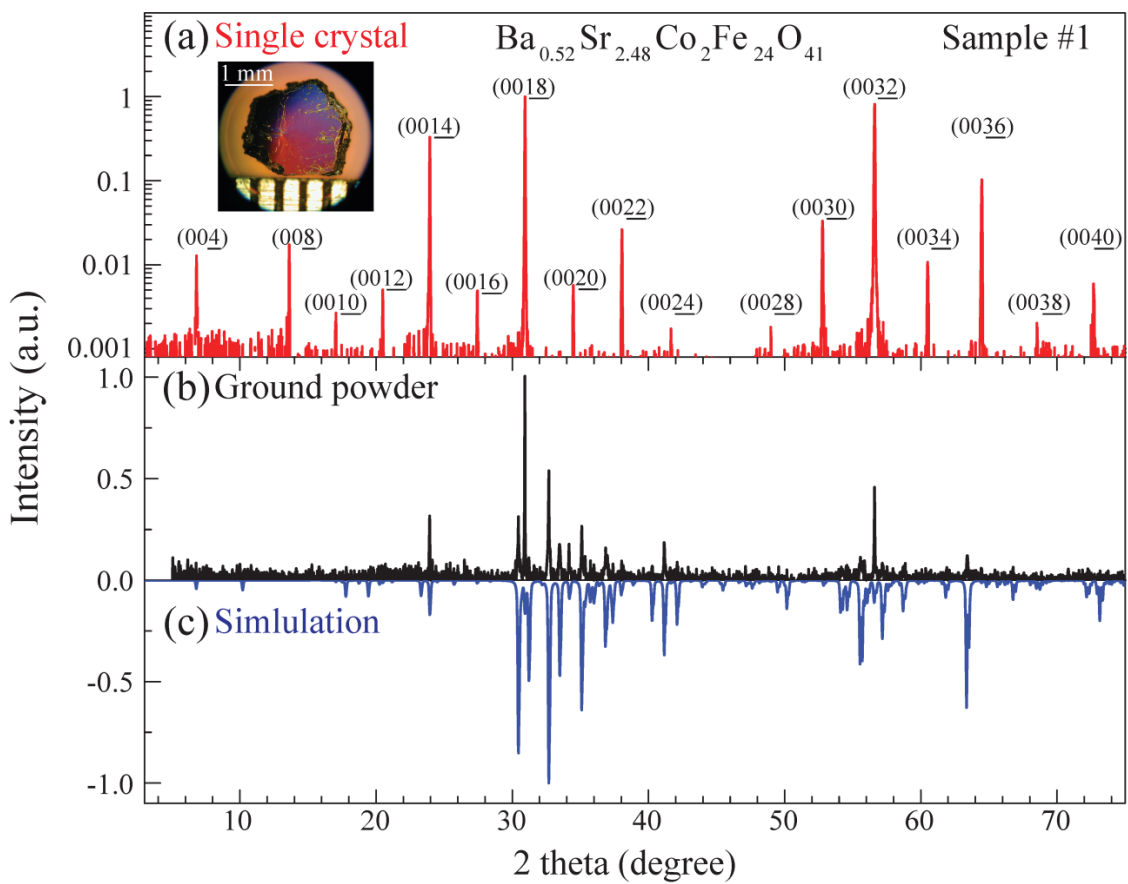

**FIG. S1** (a, b) X-ray diffraction patterns of the $Ba_{0.52}Sr_{2.48}Co_2Fe_{24}O_{41}$ single crystal and its ground powder, respectively. In (a), the scattering vector was maintained to be parallel to the $c$ axis. (c) A simulated X-ray powder diffraction pattern is also shown for the Z-type crystal with lattice constants $a = 5.867$ Å and $c = 52.00$ Å with the crystallographic point group of 6/mmm (ICSD #98328). The (0 0 $l$) peaks for the ground powder show bigger intensities than those obtained by simulation due to the preferential orientation of the crystalline grains. The inset shows the picture of a grown Z-type single crystal.

## 2. Investigations of magnetoelectric and magnetic properties

For electrical and magnetic measurements presented in the main text, two pieces of the samples were prepared from a single crystal #1. One piece with its dimension of 0.3 mm (|| [100]) × 3 mm (|| [120]) × 0.3 mm (|| [001]) was used for $P_{[100]}$ vs. $H_{[120]}$, $M_{[120]}$ vs. $E_{[100]}$, $M_{[120]}$ vs. $H_{[120]}$, and $M_{[120]}$ vs. $T$ measurements. For $M_{[001]}$ vs. $H_{[001]}$ and $M_{[001]}$ vs. $T$ measurements, another piece with the dimension 0.2 mm (|| [100]) × 0.2 mm (|| [120]) × 1 mm (|| [001]) was prepared. Both pieces, termed as Sample #1, were shaped to be a needle-like along the direction of $M$ measurements to minimize the demagnetization effects.

Temperature and magnetic field were controlled by using a Physical Property Measurement System (PPMS$^{TM}$, Quantum Design). In all the measurements, $H$ was applied along the [120] direction. For the MES (d$P$/d$H$) measurements, a customized solenoid coil was designed to generate an ac magnetic field ($H_{ac}$) of 0.02 mT at a frequency of 171 Hz, which was well compensated not to affect the superconducting magnet in the PPMS [Ref. S3]. Modulated ME charges by $H_{ac}$ were detected by a lock-in amplifier connected through a high impedance charge amplifier with a gain factor of $10^{12}$ V/C. For phase-sensitive determination of the MES signals, the phase of $H_{ac}$ was measured by a voltage pick-up coil inside the solenoid.

Both MES and magnetoelectric currents were measured and integrated to determine the change of electric polarization with magnetic fields. Before the MES and ME current measurements, a specimen went through a ME poling procedure at 305 K. For +$H_p$ & +$E_p$ poling condition (*state* **0**), the sample was kept electrically poled with +$E_p \equiv$ 230 kV/m (|| [100]) from its paraelectric state at +$\mu_0 H_p \equiv$ 2 T (|| [120]) to the magnetoelectric state at 1.2 T. Then, +$E_p$ was turned off and measurements were performed by sweeping $H$ down to -3 T or up to 3 T. Upon changing the direction of the poled electric field, we could similarly write the *state* **1** (i.e., +$H_p$ & -$E_p$). For realizing -$H_p$ & +$E_p$ poling (*state* **2**), the sample was kept electrically poled under +$E_p$ while sweeping $H$ from -2 T to -1.2 T, followed by the ME current or MES measurements from -1.2 T to -3 T or -1.2 T to 3 T at zero $E$-bias. Upon changing the direction of $E_p$, we could also realize $-H_p$ & $-E_p$ (*state* 3). For the MES measurements at different temperatures, a similar ME poling was performed at 305 K and then the specimen was cooled or warmed to the target temperature by keeping the electrical poling at 1.2 T or -1.2 T. Magnetization measurements were performed with a vibrating sample magnetometer (VSM) in the PPMS. For $M(E)$ curves, we modified a conventional VSM sample holder to allow an application of high $E$ during the $M$ measurement. For the low $H$ measurement ($|\mu_0 H| \leq 2$ mT), in particular, we carefully calibrated $H$ of the superconducting magnet in PPMS with a standard paramagnetic Pd sample [Ref. S4].

The maximum MES values at 305 K obtained for $Ba_{3-x}Sr_xCo_2Fe_{24}O_{41}$ ($2.40 \leq x \leq 3.00$) single crystals are summarized in Fig. S2. In the main text, we focused on the sample with $x = 2.48$ confirmed by EPMA (nominally, $x = 2.55$), which shows the highest value among the measured $Co_2Z$-type hexaferrites. Note that the maximum value of the $Sr_3Co_2Fe_{24}O_{41}$ single crystal is about 1150 ps/m, which is higher than the published value of 250 ps/m for a polycrystalline specimen [Ref. S5]. This observation suggests that aligned magnetoelectric domains realized in the single crystal result in higher MES value. The origin for realizing the highest MES value in the specific Sr doping needs to be understood further based on the more systematic doping dependent studies.

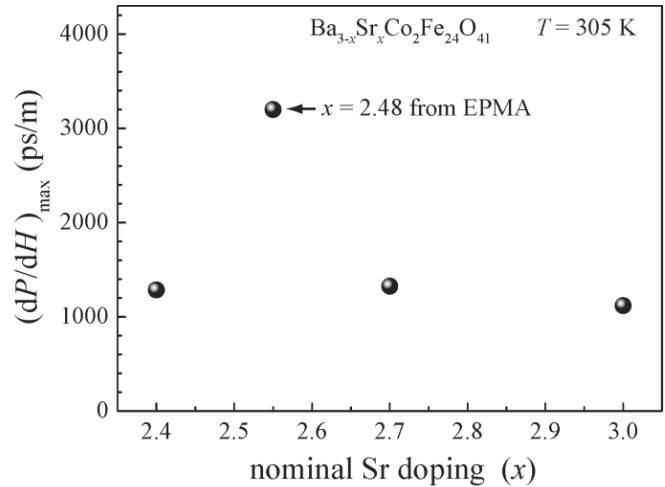

**FIG. S2** A summary of the maximum MES at 305 K for the $Ba_{3-x}Sr_xCo_2Fe_{24}O_{41}$ ($2.40 \leq x \leq 3.00$) single crystals.

## 3. Transverse conical state and development of electric polarization

The d$M$/d$T$ vs. $T$ plot obtained from the $M$ vs. $T$ curve at $\mu_0 H$ = 50 mT clearly shows a dip structure at $T_{con}$ = 410 K while the curve measured at 500 mT starts to decrease at $T_{con}$ upon cooling (Fig. 1(c) in the main text). Both of these features evidence the stabilization of the transverse conical spin order [Fig. S3]. Furthermore, we found a jump at $T_{con}$ in the specific heat vs. $T$ data, supporting the existence of a phase transition (not shown). In concurrence with the transverse conical spin order, the electric polarization develops at $T_{con}$ [Fig. S4(a)]. $\Delta P (\equiv P(H) - P(H_p \equiv 2\text{ T}))$ values at $\mu_0 H$ = 0.2 T for each temperature are determined from the isothermal $\Delta P(H)$ curves, which are obtained through the MES measurements for +$E_p$ or -$E_p$ cases. The sign of $\Delta P$ changes by reversing the direction of $E$-poling. The d$P$/d$H$ starts to appear at $T_{con}$ [Fig. S4(b)], supporting again that the electric polarization becomes finite below $T_{con}$.

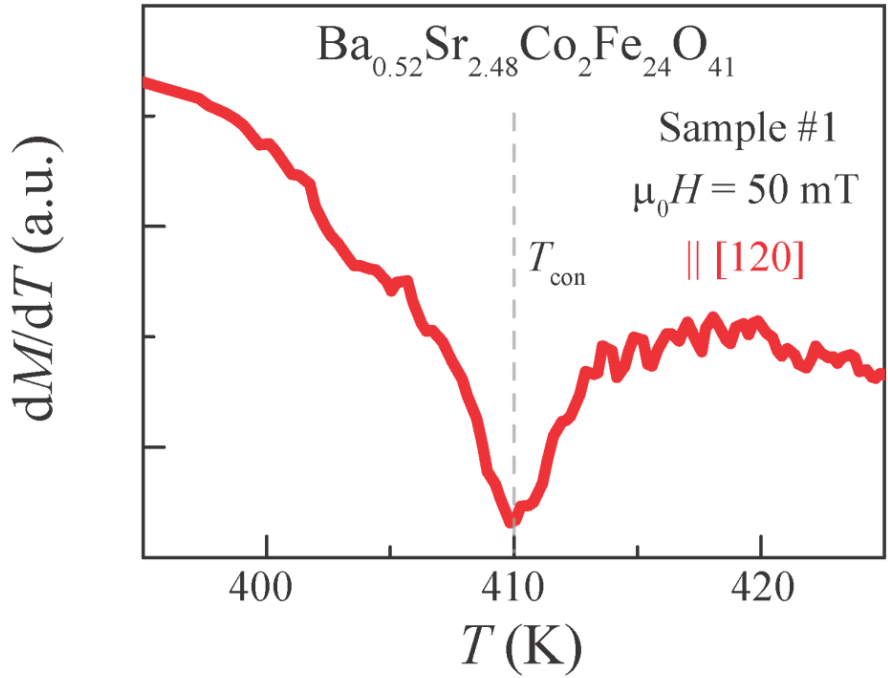

**FIG. S3** d$M$/d$T$ vs. $T$ plot estimated from $M$ (|| [120]) vs. $T$ curve at $\mu_0 H$ = 50 mT.

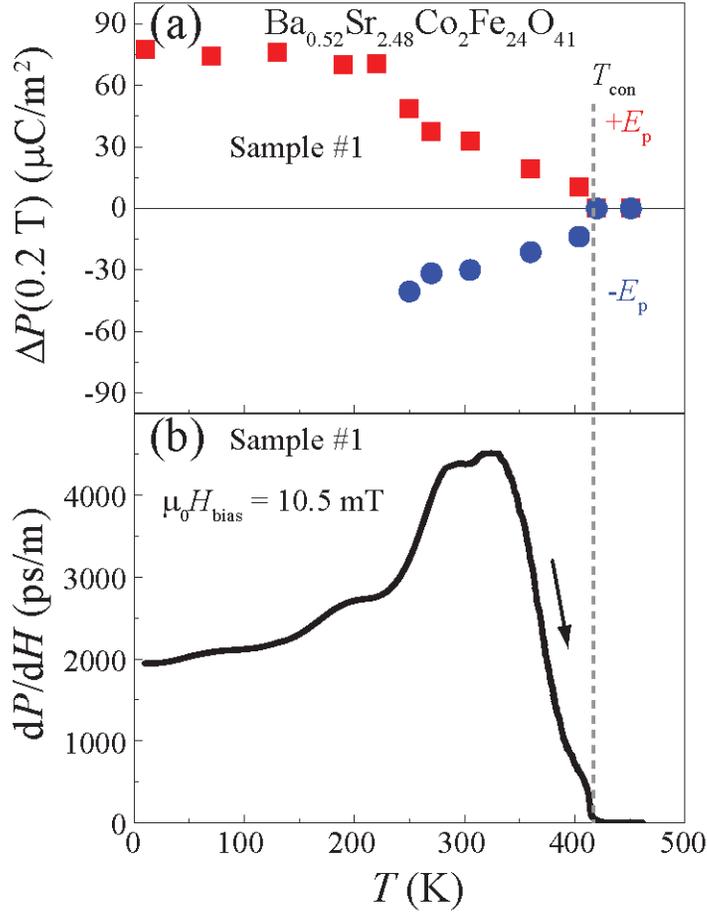

**FIG. S4** (a) Temperature-dependence of $\Delta P$ at $\mu_0 H = 0.2$ T, showing a sign change by reversing the electric poling direction. (b) Temperature-dependence of the MES at $\mu_0 H = 10.5$ mT, measured upon warming after a positive electric field poling ($+E_p$), showing a clear increase at $T_{con} = 410$ K. The curve for $-E_p$ showed a clear sign change with similar shape (not shown).

To further check whether the observation of a finite remanent $\Delta P$ is real at zero $H$-bias, we performed MES measurements after applying the post $E$-poling condition as schematically shown in Fig. S5, because a $P$-$E$ hysteresis loop at 305 K could not be obtained directly at zero $H$. Initially, the ME state "**0**" was prepared by the ME poling, $+H_p$ & $+E_p$, in which the sample was kept electrically poled with $+E_p = 230$ kV/m from the paraelectric state at $+\mu_0 H_p = 2$ T to the magnetoelectric state at 1.2 T. After then, following procedures were sequentially performed; $+E_p$

was turned off, the electrodes of the sample were electrically shorted for more than 30 minutes, and $H$ was set to zero. For a single post $E$-poling, a triangular shaped electrical pulse with a total duration of 25 sec was applied. The peak value was systematically varied, e.g., $E_{max}$ = +0.86, +0.43, 0, -0.43, and -8.86 MV/m [top, middle panel in Fig. S5]. Then, MES was measured by sweeping $H$ from 0 to 3 T. For the multiple, alternating post $E$-poling procedure, positive and negative $E_{max}$ values were applied repeatedly for more than four cycles, and lastly the same $E_{max}$ was applied, similar to the case of the single post $E$-poling [bottom, middle panel in Fig. S5].

Figures S6(a) and (b) present the evaluated $\Delta P(H)$ curves in Sample #2 for each single or multiple post $E$-poling condition. The remanent $\Delta P$ estimated from the curves [Fig. S6(a)] clearly shows a sign reversal upon changing the polarity of $E_{max}$ in the single post $E$-poling [Fig. S6(c)]. This observation strongly supports the existence of electrically polar domains even at zero $H$-bias in the $Co_2Z$-type hexaferrites. On the other hand, the non-saturating behavior of remanent $\Delta P$ values at the maximum allowed $E_{max}$ = +0.86 MV/m before electrical break-down suggests that $E_{max}$ is lower than the coercive electric field $E_c$ of possible ferroelectric domains at zero $H$-bias or the magnetoelectric domains show continuously increasing $\Delta P$ with increased $E$. As a result, the change of remanent $\Delta P$ becomes small and its sign always remains positive upon applying multiple, alternating post $E$-poling [Fig. S6(b) and (d)]. This shows that remanent $\Delta P$ becomes less susceptible to the last applied $E_{max}$ as the sample is already annealed by the multiple, alternating electric fields. Namely, after performing the multiple $E$-poling, it seems that the electrically polar domains are electrically annealed to produce more or less a positive remanent $\Delta P$.

It is also important to note that the last $E_{max}$ value still affects the remanent $\Delta P$ value. In Fig. S6(d), the remanent $\Delta P$ value, albeit positive, decreased somewhat significantly upon applying a negative $E_{max}$ as the last pulse. The application of the more negative $E_{max}$ made remanent $\Delta P$ decreased more. On the other hand, the positive remanent $\Delta P$ value remained almost same by application of a positive $E_{max}$. These observations suggest that the electrically positive polar domains, giving rise to a positive remanent Δ*P*, can be mixed with the negative domains upon applying negative electric fields while the original positive domains remain same with application of positive electric fields.

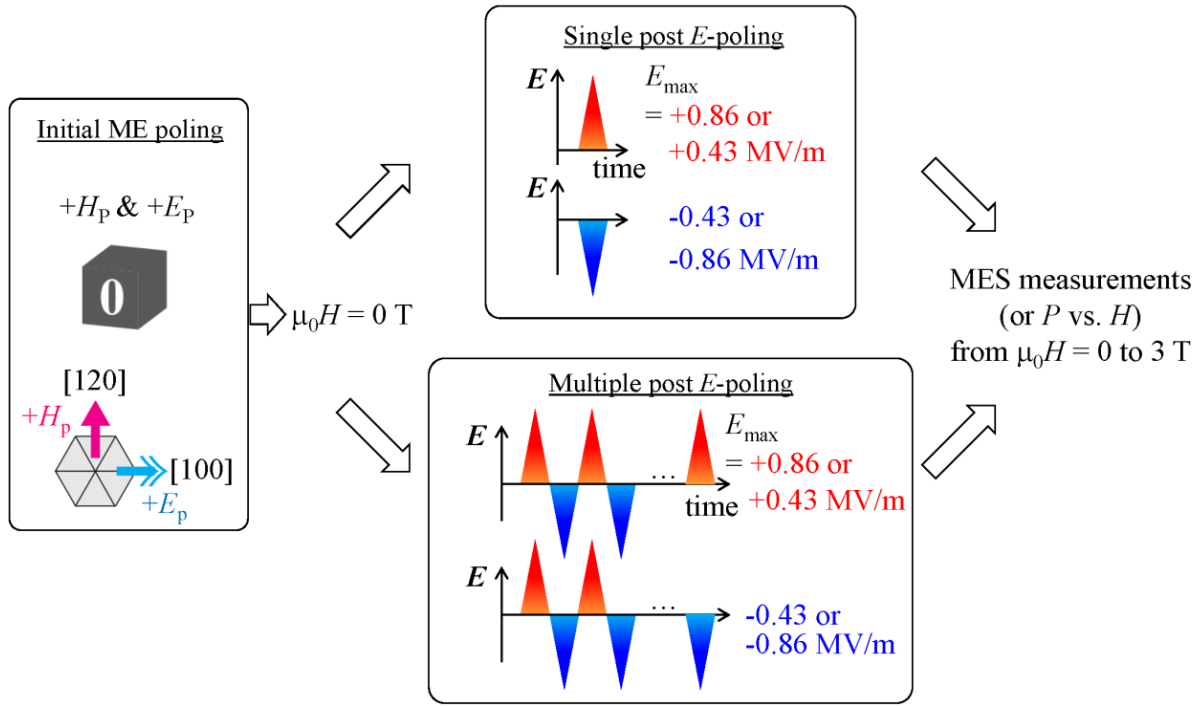

**FIG. S5** Schematic diagrams showing the post *E*-poling conditions based on a single or multiple triangular electric pulses. This additional post poling procedure was applied to test the existence of electrically polar domains in the ME state "**0**".

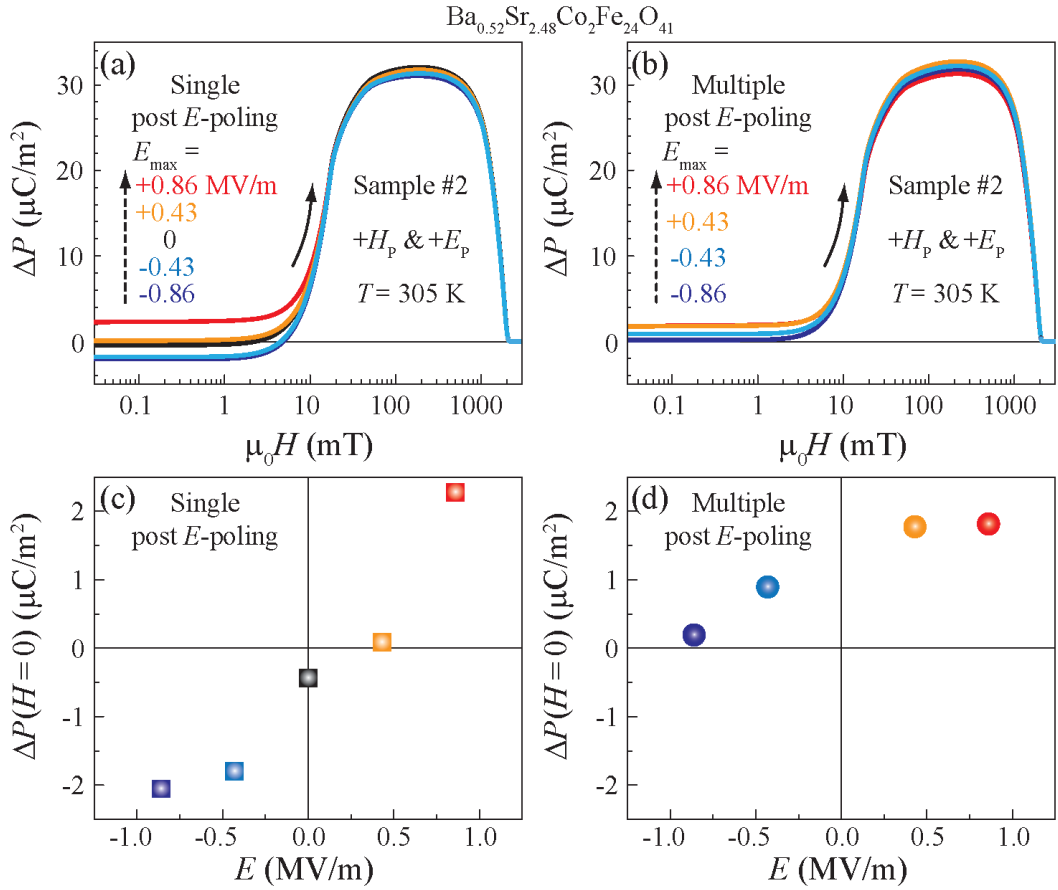

**FIG. S6** The $\Delta P(H)$ curves of Sample #2 obtained for (a) a single or (b) multiple post $E$-poling. (c), (d) $\Delta P$ estimated at zero $H$-bias after applying the post $E$-poling procedures.

## 4. Repeatable magnetization modulation by *E* at zero *H*-bias

We herein show that the modulation of magnetization is persistently repeatable for many *E*-cycles even at zero *H*-bias. In order to utilize the memory effects in the magnetization control, we first employed each ME poling, termed "*state* **0**-**3**"; *state* **0** or **1** (*state* **2** or **3**) is prepared by changing *H* from +$H_p$ (-$H_p$) to 1.2 T (-1.2 T) under application of +$E_p$ or -$E_p$ and the electric field was subsequently turned off. *M* (∥ [120]) was then measured at a fixed, bias *H* (∥ [120]) while applied *E* (∥ [100]) is swept slowly in time between -1.14 and 1.14 MV/m at a frequency of 0.02 Hz. All the curves, as demonstrated in Fig. S7(a)-(d), showed quite repeatable traces for many electric field cycles. This experimental result proves that the stored ME information is non-volatile and can be read through repeatable *M*(*E*) curves.

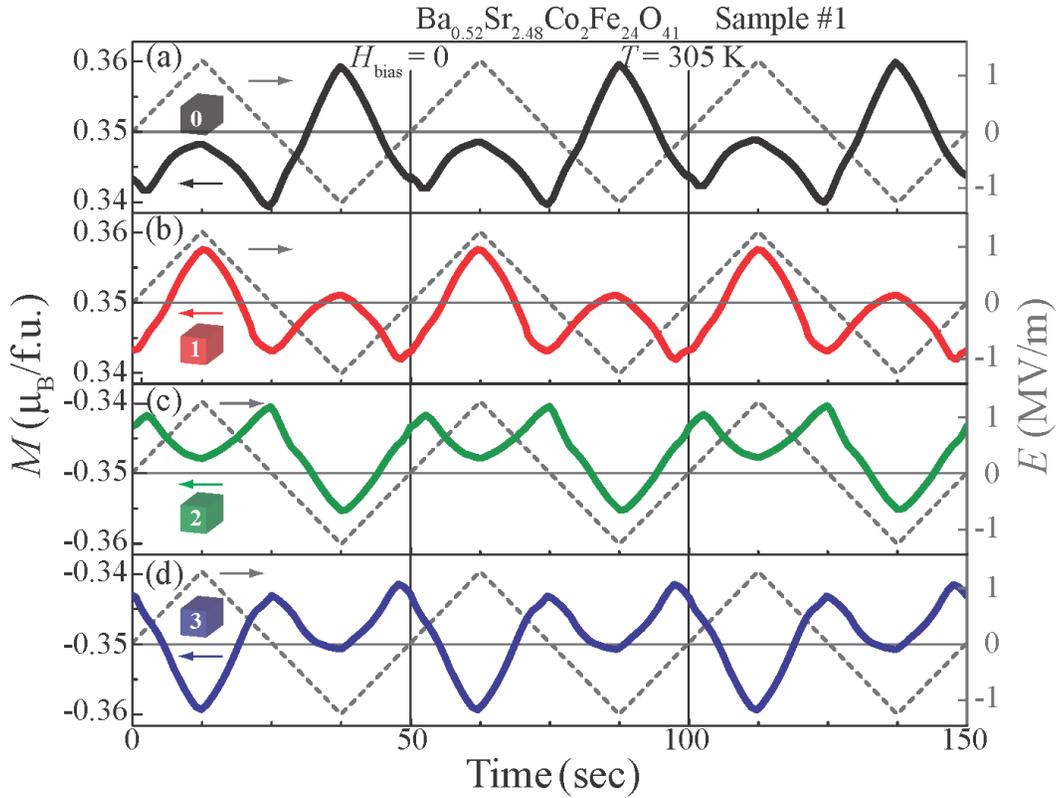

**FIG. S7** (a-d) Time dependent *M* variation for each ME state under periodically varying *E*-bias between -1.14 and 1.14 MV/m (dashed grey lines) for the four ME states (*state* **0**-**3**) in the case of Sample #1.

Furthermore, when we look into the $M(E)$ curves at zero $H$-bias in the main text [Fig. 3], the curves labeled "0" and "2" indeed show larger hysteretic part near $E = 0$, while those labeled as "1" and "3" show smaller. We postulate that such asymmetric hysteretic behaviors between the curves "0" and "1" are closely related to the ME poling and measurement history. For our $M(E)$ measurements, we first employed each ME poling, "*state* **0**-**3**". Namely, *state* **0** (**1**) is prepared by changing $H$ from +$H_p$ to 1.2 T under application of +$E_p$ (-$E_p$) and the electric field was subsequently turned off. $M$ was then measured at a fixed, bias $H$, e.g. 10.5 mT, while triangular shaped electrical pulse $E$ was applied as a sequence of 1.14 MV/m and -1.14 MV/m repeatedly. The same measurements were performed at the decreased $H$-bias from 10.5 mT, 5 mT, 1 mT, 0.5 mT, 0.4 mT, etc. Finally, $H$-bias was set to zero to obtain the results in Fig. 3. During this process, the electrically polar domains can be mostly affected by the last sequence of the former measurement. For instance, the polar domains with +Δ$P$ in the original ME states "**0**" could be more mixed with negative domains by the triangular pulse ending always in a negative $E$ to zero region. On the other hand, the polar domains with -Δ$P$ in the states "**1**", which were prepared by -$E_p$ could be less affected by this process. Therefore, the asymmetric hysteresis behavior near $E = 0$ between the curves "0" and "1" (or the curves "2" and "3") seems to be the consequence of different electric field history that was applied prior to the measurements.

To check how the history of electric field can affect the results, we performed the $M$ vs. $E$ measurements at zero $H$-bias for Sample #2 by applying a smaller $E_{max}$. When $E_{max} = 0.43$ MV/m was applied, the hysteretic part was clearly reduced [Fig. S8(a) and (b)]. These observations suggest that the mixed ME domains existing at zero $H$-bias show partial switching under high $E$ to become a main source of the hysteresis. However, it is important to note that the hysteretic part is rather small in the overall $M(E)$ curves and the characteristic asymmetry coming from the written ME information is still maintained regardless of the hysteretic region. Therefore, the free energy analysis as done in the main text becomes still possible.

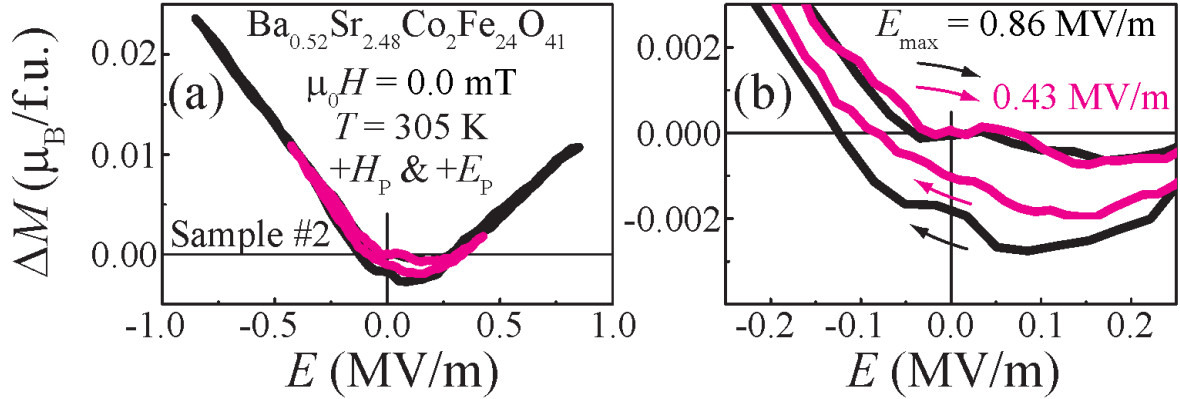

**FIG. S8** (a) $\Delta M$ (= $M(E)$ - $M(0)$) vs. $E$ curves for two different $E_{max}$ values (0.43 and 0.86 MV/m) and (b) its blown-up curve.

We also checked whether the hysteretic region is also affected by the final magnetic field at which $E_p$ was turned off. While the final magnetic field was 1.2 T in the experiment performed for Sample #1 in the main text, we herein checked the $M(E)$ curves for Sample #2 at zero $H$-bias by having the ME poling procedures ended at $H = 0$ [Fig. S9] rather than at $\pm 1.2$ T [Fig. 3 in the main text]. Indeed, we could reproduce in Fig. S9 all the main features observed in Fig. 3. Namely, the four, independent $M(E)$ curves, showing characteristic asymmetry for each ME states, could be successfully reproduced. Moreover, the overall hysteretic regions in the four $M(E)$ curves were reduced as compared with those shown in Fig. 3. However, it is also important to note that the overall characteristic shape is still maintained for the $M(E)$ curves both in Fig. 3 and Fig. S9. Therefore, we conclude that the final magnetic field at which $\pm E_p$ was switched off does not change the main physics relevant to the free energy analyses either. We plan to separately report the systematic free energy analyses results for each ME poling state as well as $E_{max}$.

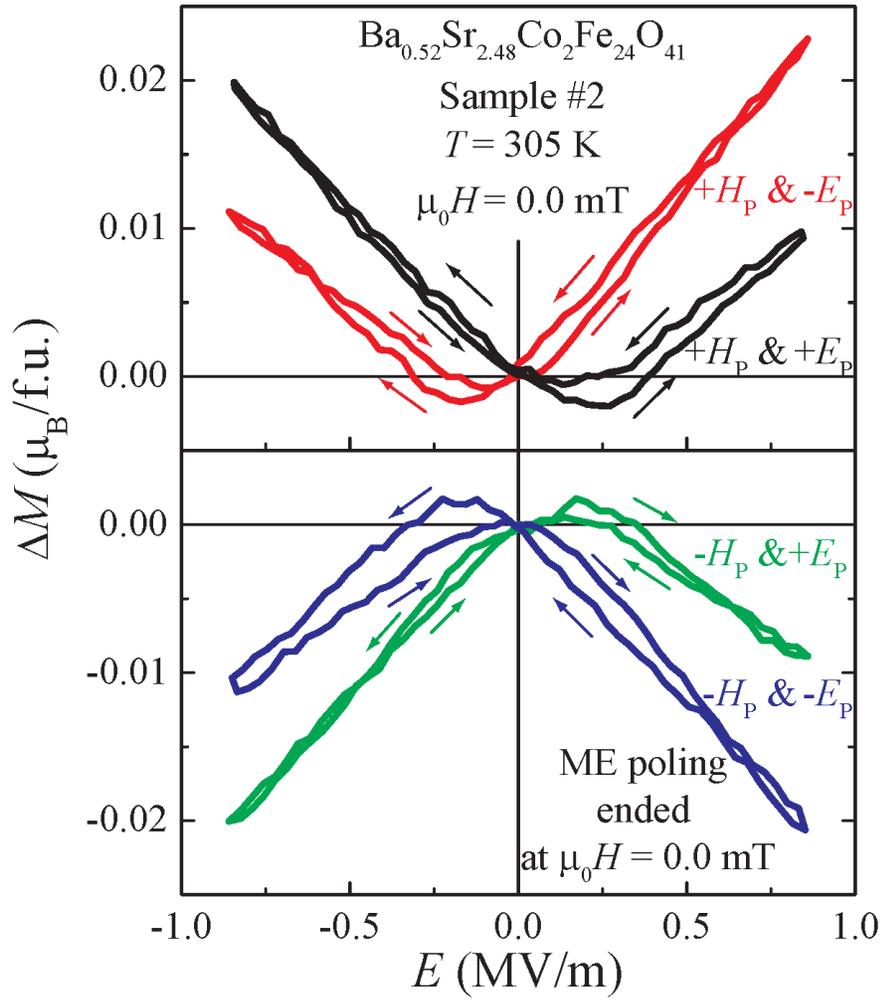

**FIG. S9** $\Delta M$ (= $M(E) - M(0)$) vs. $E$ curves of Sample #2 at zero $H$-bias for four different ME poling states for which $E_p$ was maintained down to $H = 0$.

## 5. Magnetoelectric coefficients

The ME coefficients α, β, γ and π can be extracted from the *M(E)* curves (*E*-decreasing run) near zero *H*-bias for each different ME poling condition (Fig. 4(a)-(d) in the main text), as summarized in Table S1.

| ME poling | α (ps/m) | β (ps/m·mT) | γ (ps/MV) | π (ps/MV·mT) |
|---|---|---|---|---|
| *state* **0** (+$H_p$ & +$E_p$) | -67 | 467 | 168 | 584 |
| *state* **1** (+$H_p$ & -$E_p$) | 67 | -622 | 159 | 734 |
| *state* **2** (-$H_p$ & +$E_p$) | 67 | 575 | -118 | 614 |
| *state* **3** (-$H_p$ & -$E_p$) | -67 | -700 | -180 | 608 |

**TABLE S1** ME poling dependence of ME coefficients.

The crystallographic point group of $Co_2Z$-type hexaferrite is known as 6/mmm. The exact magnetic point group is not known yet at every magnetic field. However, if we suppose the transverse conical structure as depicted in Fig. 1(b) [Ref. S5], which is known to be realized at a low *H* (|| [120]) region, the allowed magnetic symmetry corresponds to 2'. According to Ref. S6 and S7, the magnetic point group 2' can have both non-trivial linear and quadratic ME tensors as $P_{[100]} = \alpha_{12}H_{[120]} + \beta_{122}H_{[120]}^2 + ...$ and $M_{[120]} = \alpha_{21}E_{[100]} + \gamma_{211}E_{[100]}^2 + ...$ . The existence of such finite tensor components is indeed quite consistent with our experimental observations to find the non-zero α, β, and γ values. The correspondence between the coefficient π for the higher order magnetoelectric effect and the macroscopic tensor components has not been yet established to our knowledge [Ref. S6]. However, an ambiguity still exists for the proper magnetic point group at zero *H* if the spin configuration cannot be maintained as in Fig. 1(b), possibly due to mixing of magnetic and ME domains. At the present stage, one needs an independent study to confirm a proper spin configuration at zero *H* either by using a neutron scattering or an imaging tool.

Although we have mainly extracted the ME coefficients from the $M(E)$ curves in this study, we can directly obtain the ME coefficients from the MES ($dP/dH$) measurements as well. Starting from $F(E, H)$ in the main text, minimization of $F$ with respect to $E$ result in the $P(E, H)$ form as followings:

$$P(E, H) = \varepsilon_0 \varepsilon E + \alpha H + \frac{1}{2}\beta H^2 + 2\gamma EH + \pi EH^2 + ...$$

and then the direct MES for the case of $E = 0$ can be extracted as

$$\frac{dP}{dH}(E = 0, H) = \alpha + \beta H + ... \quad .$$

The experimental data of $dP/dH$ near zero $H$-bias for each ME poling clearly showed the linear behavior as expected in the above equation and could be thus well fitted with $\alpha + \beta H$, as demonstrated in Fig. S10. Through this procedure, a new set of $\alpha$ and $\beta$ were obtained as in Table S2.

Careful investigation of Table S1 reveals that the ME coefficients, $\alpha$, $\beta$, and $\gamma$, obtained from the $M(E)$ curves change signs upon changing the written ME states, satisfying the empirical relationship, $\alpha(\pm H_p)(\pm E_p) < 0$, $\beta(\pm E_p) > 0$ and $\gamma(\pm H_p) < 0$. Furthermore, from the independent, direct MES ($dP/dH$) measurements for each ME poling [Table S2], we also find that the ME coefficients $\alpha$ and $\beta$ show similar sign changes. All these experimental results show that the four different ME states can be represented by different set of ME coefficients, and both direct ME and converse ME effects at 305 K can appear in accordance with those stored ME states, characterized by a set of the ME coefficients.

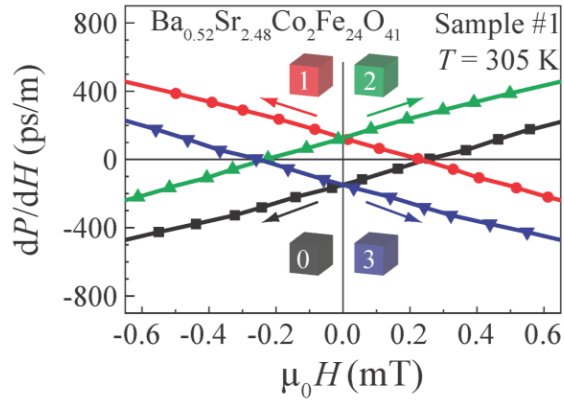

**FIG. S10** d$P$/d$H$ near zero $H$ for four different ME poling conditions.

| ME poling | α (ps/m) | β (ps/m·mT) |
|---|---|---|
| state **0** (+$H_p$ & +$E_p$) | -149 | 575 |
| state **1** (+$H_p$ & -$E_p$) | 124 | -539 |
| state **2** (-$H_p$ & +$E_p$) | 124 | 532 |
| state **3** (-$H_p$ & -$E_p$) | -149 | -591 |

**TABLE S2** ME poling dependence of ME coefficients extracted from d$P$/d$H$ curves in Fig. S10.